\documentstyle[11pt,moriond,epsfig]{article}
\begin{document}
\vspace*{4cm}
\title{Galactic Free-free Emission and H$\alpha$}
\author{J.G. Bartlett$^1$ \& P. Amram$^2$}
\address{$^1$Observatoire de Strasbourg, 11 rue de l'Universit\'e, 
	67000 Strasbourg, France\\
	$^2$IGRAP, Observatoire de Marseille, 2 Place Le Verrier,
	13248 Marseille Cedex 4, France} 
\maketitle
\abstracts{We review present understanding of
Galactic free--free emission and its possible importance to CMB fluctuation
measurements. Current results, from both ``direct'' observations in the
microwave band and from H$\alpha$ studies, suggest that this foreground does
not represent a serious obstacle to mapping the CMB; however, this is based
on limited information and we emphasize the need for more exhaustive
studies. We also present some preliminary results based on our recent H$%
\alpha$ observations near the South Pole CMB data sets. The fluctuation
amplitude seen in H$\alpha$ indicates that the detected CMB fluctuations are
not significantly contaminated by free--free emission, at least if the
diffuse gas is at a temperature of $T\sim 10^4$ K.}

\section{Galactic Free--free Emission}

The three main sources of microwave emission from our Galaxy are synchrotron
radiation, thermal dust emission and bremsstrahlung (free-free emission).
Among these three, bremsstrahlung poses the greatest difficulty for cosmic
microwave background (CMB) observations for two reasons: 1/ it dominates
over the frequency range where the total Galactic emission is minimal, and
hence where the majority of CMB experiments are performed, and 2/ it
dominates {\em only} over a small range of frequencies (see Fig. 3 of
Kogut et al. 1996a). Thus, it is the principal Galactic contaminant of many
CMB experiments, but it cannot be traced by observing at either much higher
or lower frequencies. For example, the dust and synchrotron foregrounds may
be mapped in the infrared and radio, respectively. Such maps prove very
useful for foreground control and elimination. There exists, nevertheless, a
tracer of Galactic bremsstrahlung -- Hydrogen H$\alpha $ in emission, which
is produced by the same warm ionized medium (WIM) responsible for the
bremsstrahlung. Unfortunately, there is at present no full--sky map
available in H$\alpha $. In what follows, we review what is known about
Galactic free--free emission, with an emphasis on implications for CMB
experiments. We will first consider ``direct'' observations of the
foreground in the microwave band (from CMB work) and then consider in detail
studies based on the H$\alpha $ line. An excellent recent review of the
subject has been given by Smoot (1998). To focus the discussion, we will
center it on two primary questions: 1/ How well correlated are dust and
free--free emissions? 2/ Might there be a hot component of the interstellar
medium (ISM) producing Galactic free--free emission which could violate
limits based on a simple interpretation of current H$\alpha $
observations?

\subsection{Direct Observations of Bremsstrahlung}

A summary of direct observations of Galactic bremsstrahlung in terms of  
equivalent brightness temperature variations at 40 GHz (Q--band)
-- $\Delta T_{ff}$ -- is
given in Table 1. All these results come from CMB experiments. The three DMR
maps, obtained at 31.5, 53 and 90 GHz, have been combined to optimize a signal
with the spectral dependence of free--free emission, $\sim \nu^{-2}$ (Kogut
et al. 1996a, 1996b) (this has been done for all three Galactic emissions;
the maps are available at {\tt http://nssdca.gsfc.nasa.gov/astro/cobe/cobe%
\_home.html}). Unfortunately, the signal--to--noise in the resulting
free--free map is too low to follow in detail the distribution of Galactic
bremsstrahlung. Nevertheless, based on the COBE 4 year results, 
Kogut et al. (1996b)
constrain the brightness fluctuations due to bremsstrahlung to $\Delta 
T_{ff} = 9
\pm 7 \mu$K. This provides a 95\% upper limit of $\Delta T_{ff} < 23 \mu$K 
on the 
{\em total} Galactic free--free emission. For comparison, the extracted
CMB fluctuation amplitude (brightness variations at $10^o$) 
is $29 \pm 1 \mu$K (Banday et al. 1997).

In order to boost the signal--to--noise, Kogut et al. (1996b) also searched
for galactic emission by cross--correlating the three DMR and the DIRBE 140
micron maps, finding a significant result. The spectral characteristics of
the correlated signal are those of a combination of dust and free--free
emission, with the free--free contributing $\Delta T_{ff} = 12 \pm 5 \mu$K. 
Within the errors, this is
consistent with the above upper limit, but it does leave room for a
non--correlated component of the same amplitude. \newline

\begin{tabular}{|c||c|c|c|c|}
\multicolumn{5}{c}{{\bf {\large Table 1: Summary of Direct Free--Free
Constraints}}} \\ 
\multicolumn{5}{c}{} \\ \hline\hline
reference & resolution & patch size & $\Delta T_{ff}$ at 40 GHz & comments \\ 
\hline\hline
Kogut et al. 1996b: &  &  &  &  \\ 
(DMR/DIRBE) & $10^o$ & $|b|>20^o$ & $12\pm 5$ $\mu$K & dust correlated \\ 
(DMR) &  &  & $9\pm 7$ $\mu$K & total emission \\ \hline
de Oliveira--Costa et al. 1997 & $1^o$ & $7.5^o\times 7.5^o$ & $17.5\pm 9.5$ 
$\mu$K & dust correlated \\ 
(Saskatoon) &  &  &  & no spectral info. \\ \hline
Leitch et al. 1997 & $0.3^o$ & $DEC=88^o$ & $\sim 27$ $\mu$K & total emission
\\ 
(OVRO 5.5 \& 40 m) &  &  &  &  \\ \hline\hline
\end{tabular}
\newline
\vspace{0.5cm}

	Oliveira--Costa et al. (1997) performed a similar analysis on
the Saskatoon data, cross--correlating the CMB results with the DIRBE far
infrared maps. They marginally detect a non--zero correlation with an
amplitude consistent with that found by Kogut et al. (1996b), extending the
free--free/dust correlation down to angular scales of $\sim 1$$^{\circ}$ 
(the
COBE results are restricted to $>$7$^{o}$). These results raise the first
question posed in the introduction: Just how well does dust emission trace
Galactic bremsstrahlung? The COBE and Saskatoon analyses demonstrate the
existence of a free--free component correlated with Galactic dust emission,
but it should be remembered that the possibility of an equally important,
non--correlated component remains. The angular power spectrum of the
dust--correlated component is $P(l)\sim l^{-3}$ (Kogut et al. 1996b),
showing a rapid fall--off towards small angular scales. This power spectrum
is not at all surprising -- it is simply the power spectrum already observed
for the dust distribution (Gautier et al. 1992; Wright 1998). 
It is worth noting that
such a power spectrum implies that the true sky variance remains roughly constant towards small scales, but
that estimates of this variance in restricted patches of sky 
may be ``pulled'' far from the true variance by the important 
correlations on scales equal to the patch size; in other words, 
the uncertainty on the estimate will be larger than that deduced
by assuming $N_{\rm pixel}$ independent measurements of the sky
brightness.  This should be kept 
in mind when interpreting the estimates given in
the tables.

Complicating matters somewhat is a recent result reported by Leitch et al.
(1997). Using the OVRO 5.5 and 40 meter telescopes at frequencies of 32 and
14.5 GHz, they find a large signal in a ring at constant declination around
the North Celestial Pole (Saskatoon region) with the spectral
characteristics of free--free emission. The corresponding 40 GHz amplitude
on an angular scale of $0^o.3$ is $\Delta T_{ff} \sim 27 \mu$K, slightly larger
than the upper limits imposed on larger scales by the previous observations
or by the extrapolation of the dust power spectrum. This result could
indicate the emergence on smaller scales of an important bremsstrahlung
component uncorrelated with dust. We return to this point below when
discussing the H$\alpha$ results.

\subsection{H$\alpha$ Studies}

The ionized gas responsible for the free--free emission is also a source of
line emission, in particular Hydrogen H$\alpha$. This optical line, at 6563
angstroms, may thus be used to trace the gas distribution and, hence, map
the free--free foreground (Ly$\alpha$ would do the same, but absorption is a
serious problem). In the following, we present simplified versions of the
various formulae; a more detailed treatment can be found in Valls--Gabaud
(1998). The H$\alpha$ intensity, usually expressed in terms of Rayleighs (R$%
=2.41\times 10^{-7}$ergs/cm$^2$/s/ster), is given by 
\begin{equation}  \label{Ia}
I_\alpha = (0.36\;\mbox{R}) \left(\frac{EM}{\mbox{cm$^{-6}$ pc}}\right)
T_4^{-0.9}
\end{equation}
where $T_4\equiv T/(10^4$ K) and $EM$ represents the emission measure. This
expression is valid for temperatures $T_4\le 2.6$ (Leitch et al. 1997).
Free--free emission depends on the same quantities (given here for pure
Hydrogen and in the limit as $h\nu/kT\rightarrow 0$): 
\begin{equation}  \label{ff}
T_{ff} = \frac{(5.43\; \mu\mbox{K})}{\nu_{10}^2 T_4^{1/2}} \left(\frac{EM}{%
\mbox{cm$^{-6}$ pc}}\right) g_{ff}
\end{equation}
where the observation frequency is $\nu = \nu_{10}10^{10}$ GHz and g$_{ff}$
is the thermally averaged gaunt factor: 
\begin{equation}
g_{ff} = 4.69(1+0.176\ln T_4-0.118\ln \nu_{10})
\end{equation}
The relation between the line and free--free emissions is particularly
simple due to the fact that both depend primarily on the emission measure.
Combining Eqs. (\ref{Ia})and (\ref{ff}) yields 
\begin{equation}  \label{TffIa}
T_{ff} = (15\;\mu\mbox{K}) g_{ff} T_4^{0.4} \nu_{10}^{-2} \left(\frac{%
I_\alpha}{\mbox{R}}\right)
\end{equation}
Note that this is an expression for the absolute free--free intensity,
$T_{ff}$, to be distinguished from its variation, $\Delta T_{ff}$.
\newline

\begin{tabular}{|c||c|c|c|c|}
\multicolumn{5}{c}{{\bf {\large Table 2: H$\alpha$ Results}}} \\ 
\multicolumn{5}{c}{} \\ \hline\hline
reference & resolution & patch size & $\Delta T_{ff}$ at 40 GHz & comments \\ 
\hline\hline
Reynolds et al. 1990 & $0^o.8$ & diverse pointings & $\sim 10$ $\mu$K & 
Fabry--Perot \\ 
Reynolds et al. 1992 & $0^o.8$ & $12^o\times 10^o$ & $\sim 10$ $\mu$K & 
Fabry--Perot \\ \hline
Gaustad et al. 1996 & $0^o.1-1^o$ & $7^o\times 7^o$ & $<5$ $\mu$K & 
broad--band \\ \hline
Simonetti et al. 1996 & $1^o$ & diam $=30^o$ & $<5$ $\mu$K & broad--band \\ 
\hline
Marcelin et al. 1998 & $0^o.5$ & 24 pointings & $<5$ $\mu$K & Fabry--Perot
\\ \hline\hline
\end{tabular}
\newline
\vspace{0.5cm}

The ideal would therefore be to have a sensitive all--sky map in H$\alpha$
to control the free--free foreground; this does not at present exist. For
many years, R. Reynolds has been carrying out H$\alpha$ observations of the
Galaxy with a double Fabry--Perot spectrometer (needed to separate Galactic 
$H\alpha$ from the much larger geocoronal H$\alpha$ emission). A good review
of these results and a summary of what is known about the distribution of
the WIM is given by Reynolds (1990). For the essential, the gas seems to be
distributed above and below the Galactic Plane with a characteristic scale
height of $\sim 1$ kpc. It is thought that the gas has a temperature close
to $T_4=1$. These conclusions are the result of diverse, pointed
observations (a rough picture of the gas distribution is given in Reynolds
1990). A significant improvement will soon come from the WhaM survey
(Reynolds et al. 1998), an effort to map the northern sky in H$\alpha$ at a
resolution of 1$^{\circ}$ (the WhaM web page is a very useful source of
information -- {\tt http://www.astro.wisc.edu/wham/WhaM.html}).

A rough summary of current results from H$\alpha$ is given in Table 2, 
where we have assumed $T_4=1$.  We note that the small--scale 
observations are essentially upper limits, but that Reynolds
actually sees fluctuations at $\sim 2$ R.  From this compilation, 
it would seem that the
variations on small scales are not very large and should not pose a problem
for CMB experiments, which are expecting signals in the range $30-100 \mu$K.
An important caveat is always that, up till now, these observations cover a
rather small fraction of sky.  The last entry, Marcelin et al. (1998),
is discussed below.    

The Gaustad et al. (1996) and Simonetti et al. (1996) 
observations cover the Saskatoon area (North Celestial Pole).  
They are based narrow
band filters instead of the high resolution spectrographs used by 
Reynolds et al. and Marcelin et al.  
This means that they are unable to extract
the geocoronal from the Galactic signal, implying that only
upper limits on $H\alpha$ fluctuations can be deduced. Their strict limits 
argue against a significant level of
contamination in the CMB results from Saskatoon. However, as already
mentioned, Leitch et al. (1997) report a detection of a much stronger signal 
{\em within the same region}. As pointed out by Leitch et al., this
discrepancy could be explained by the presence of a hot component of the
ISM, say around $T\sim 10^6$ K (see Eq. [\ref{TffIa}]), bringing up the
second question posed in the introduction. It is important to realize that
this temperature corresponds to the virial temperature of the Galactic halo;
thus, we could be seeing some evidence for a hot medium in the halo of our
Galaxy, or perhaps in the Local Group. This possibility may be constrained
by X--ray observations, as discussed by Smoot (1998).

Attempting to clarify the nature of the dust/free--free relation, Kogut
(1997) compared the Reynolds et al. (1992) patch with the DIRBE maps and
found a significant correlation. This is the best way to try and answer our
first question concerning the dust/free--free correlation, because the H$%
\alpha$ gives information on the total free--free emission. Looking at
Kogut's figure 2, it is clear that there remains a significant variation
around the regression line. This could represent a non--correlated component
of bremsstrahlung contributing a signal perhaps as large as the correlated
component. If so, dust emission would not be sufficient to trace and remove
all of the free--free foreground.

\section{A New Fabry-Perot H$\alpha$ Study}

In this section, we present our recent observations in H$\alpha $ taken near
the South Pole CMB data sets. For a complete description of the instrument,
see Amram et al. (1991).

\subsection{Observations}

The observations were made from the ESO site at La Silla (Chile) in November
1996 with a 36 cm telescope equipped with a scanning Fabry--Perot
interferometer. The field of view is $38' \times 38'$, the spectral sampling
step was either 5~km~${\rm s^{-1}}$ or 2.5~km~${\rm s^{-1}}$ (i.e.
0.10~\AA~or 0.05~\AA), depending on the scanning process adopted (24 or 48
channels over the free spectral range of 115~km~${\rm s^{-1}}$, i.e.
2.5~\AA, of the Fabry--Perot interferometer). The
lines passing through the 8~\AA\ FWHM filter were:
the Galactic H$\alpha$ line we are
looking for, the geocoronal H$\alpha$ emission and the OH night--sky line at
6568.78~\AA. These two parasitic lines are brighter than the Galactic H$%
\alpha$ line we are looking for, the geocoronal line being typically twice
as bright as the OH line and 10 times brighter than the Galactic line. In
order to compare the Galactic H$\alpha$ emission fluctuations with the South
Pole CMB results (Schuster et al. 1993; Gundersen et al. 1995), we selected
fields at declinations of $-63^{\circ}$ and $-62^{\circ}$. Our fields were
separated by 15 mn in right ascension, which is about $1^{\circ}45'$ on the
sky, thus offering a fair coverage of each exposed band. We observed 19
fields at $-62^{\circ}$ (from $\alpha$ = 23${\rm ^{h}}$50 to $\alpha$ = 4$%
{\rm ^{h}}$20) and 5 fields at $-63^{\circ}$ (from $\alpha$ = 1${\rm ^{h}}$%
35 to $\alpha$ = 2${\rm ^{h}}$50). Some of these fields were observed twice,
on different nights, to check the reproducibility of our measurements.  The
observing conditions were fairly good, and a standard 2${\rm ^{h}}$ exposure
time was adopted.

\subsection{Data reduction}

Due to the ``wrap--around'' nature of the FSR (Free Spectral Range) of the
Fabry-Perot interferometer (FSR=2.52~\AA), the OH night sky line at
6568.78~\AA~appears closer to the H$\alpha$ lines (geocoronal and Galactic)
than it actually is, with an apparent separation of only 1~\AA~with respect
to H$\alpha_{o}$ (see Fig. 1). The motion of the earth around the sun and
the motion of the sun in the galaxy were combined in such a manner that the
separation between the Galactic and geocoronal H$\alpha$ lines remained
approximately constant, around 0.5~\AA, along the bands of sky observed in
November. The extraction of Galactic H$\alpha$ emission is not easy,
since it is typically 10 times fainter than the parasitic lines, whose width
(FWHM around 0.35\AA) and shape (not far from gaussian) also tend to bury
the Galactic signal in their wings. To improve the signal--to--noise ratio
and the spectral resolution, we selected a $30'$ diameter disk centered on the
interference rings observed in each field. The H$\alpha$ emission profile
obtained for each observed field is thus the addition of the profiles of all
pixels (about \mbox{31000}) found within $15'$ from the center
of the field.

The OH night--sky line at 6568.78~\AA~is in fact the sum of two close
components of the same intensity, one at 6568.77~\AA~and the other at 
6568.78~\AA.
More complicated is the case of the geocoronal emission line, with not less
than seven fine structure transitions. The two main components, produced by
Lyman $\beta$ resonance excitation, are found at 6562.73~\AA~and
6562.78~\AA, with 2:1 ratio (Yelle \& Roesler 1985). However, these
two components were insufficient when we decomposed our observed profiles, a
residual remaining systematically at 6562.92~\AA. This is due to cascade
excitation which is particularly strong for the 7th component at
6562.92~\AA~(Nossal 1994). Although the cascade contribution is not
accurately known, it proved satisfactory to use the Meier Model cited in
Nossal's thesis, adding a component at 6562.92~\AA~with a 1:6 ratio compared
with the brightest component at 6562.73~\AA.

\begin{figure}[tbp]
\epsfclipon
\psfig{figure=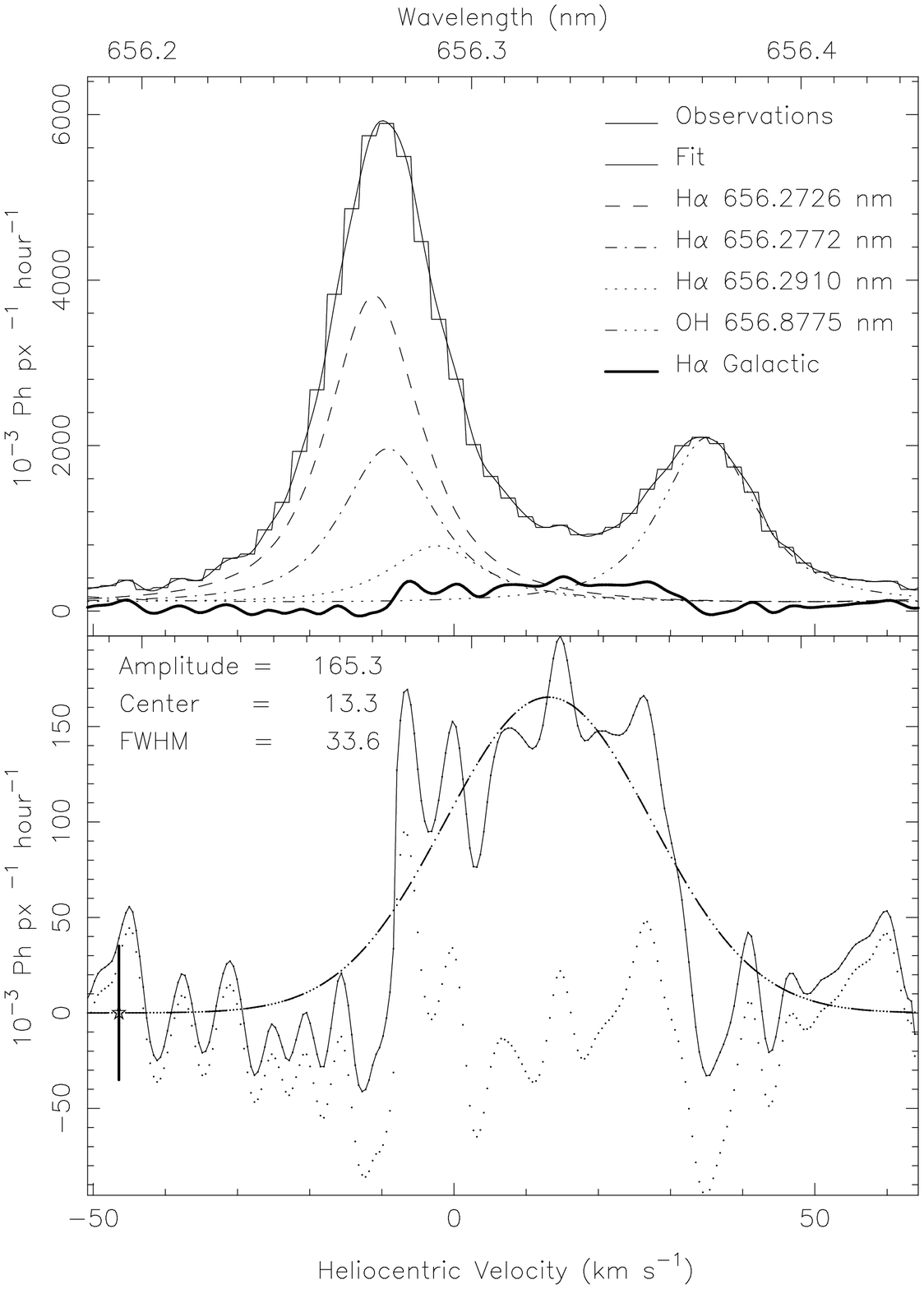,height=4.2in}
\psfig{figure=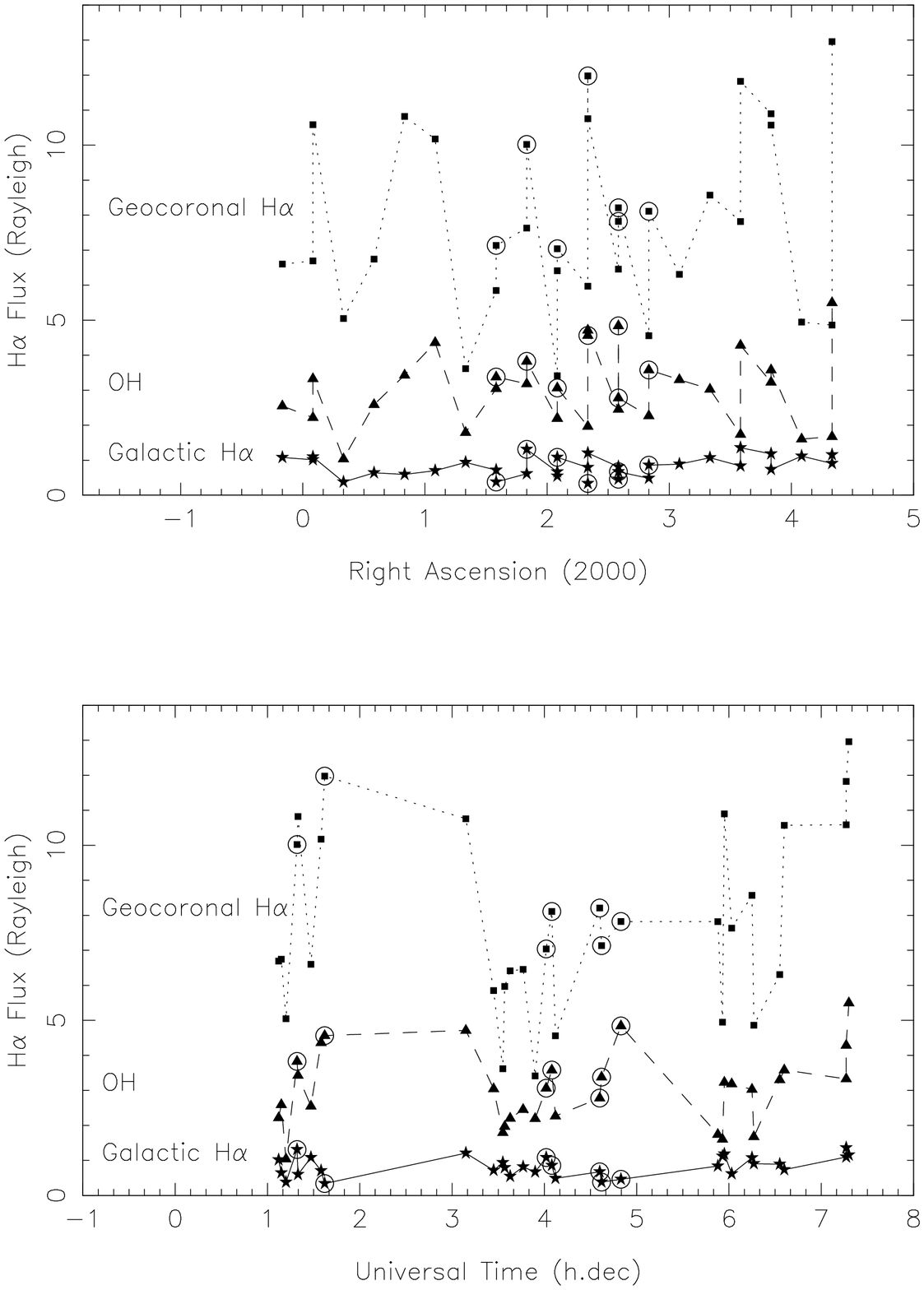,height=4.2in}
\caption{
LEFT: -Top- Example of the decomposition of the observed line
profile into geocoronal H$\alpha$ emission (3 components) and OH night--sky
line. The remaining signal (thick line) shows the Galactic H$\alpha$
emission. -Bottom- Best fit of the same Galactic H$\alpha$ emission,
amplified 20 times, by a gaussian. The shift with respect to 0 km s$^{-1}
$ heliocentric velocity is explained by the motion of the sun with respect
to the LSR. 
RIGHT:  Geocoronal  H$\alpha$ (squares + dotted), OH (triangles +
dashed) and Galactic H$\alpha$ (stars + solid) line intensities as 
a function of -Top- right ascension (2000); 
-Bottom- 
mean observing time (UT). The effect of the solar depression 
angle is visible.
Circled symbols represent the fields at a declination -63$^{\circ}$;
the other fields are at declination -62$^{\circ}$.
}
\end{figure}

After subtraction of the night--sky lines, a residual was found at the
expected velocity for Galactic H$\alpha$ emission, that is to say, around
zero in V$_{LSR}$ (radial velocity in the local standard of rest), and with
the expected width, around 35~km~${\rm s^{-1}}$, according to Reynolds
(1990). Fig. 1 shows an example of a profile decomposition for one of our
fields, together with the gaussian fit of the Galactic emission below. The
width of the gaussian was left as a free parameter and adjusted
automatically for the best fit. This width was typically found to vary
between 25 an 50~km~${\rm s^{-1}}$. The absolute calibration in intensity 
was made by observing the HII region N11E in the Large
Magellanic Cloud (Caplan \& Deharveng 1985).

\subsection{Results}

We find that the Galactic H$\alpha$ emission along the observed bands, at $%
-62^{\circ}$ and $-63^{\circ}$, varies between 0.2 R and 1.4 R, in agreement
with intensity values measured by Reynolds (1990) in the northern hemisphere
far from the galactic plane. Fig. 2 shows the measured intensity for the
Galactic H$\alpha$ emission along the two declination bands.  
The error bars are the average {\em rms}
difference between the signal and the fitted gaussian, found to be 0.35
Rayleigh. Reynolds (1990) gives intensities of the Galactic H$\alpha$
emission with an uncertainty of 0.4 Rayleigh.

\begin{figure}[tbp]
\epsfclipon\psfig{figure=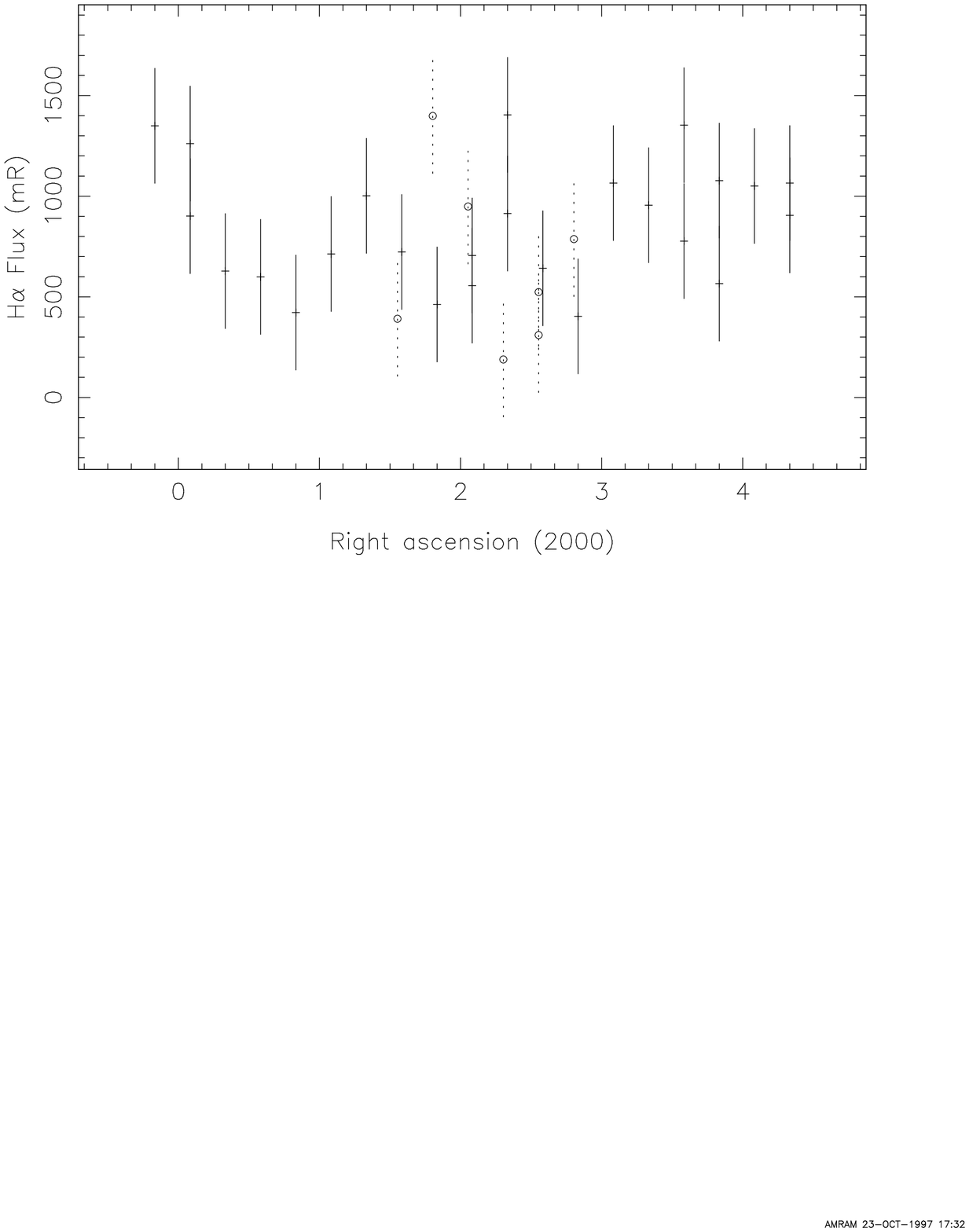,height=2.0in} %
\psfig{figure=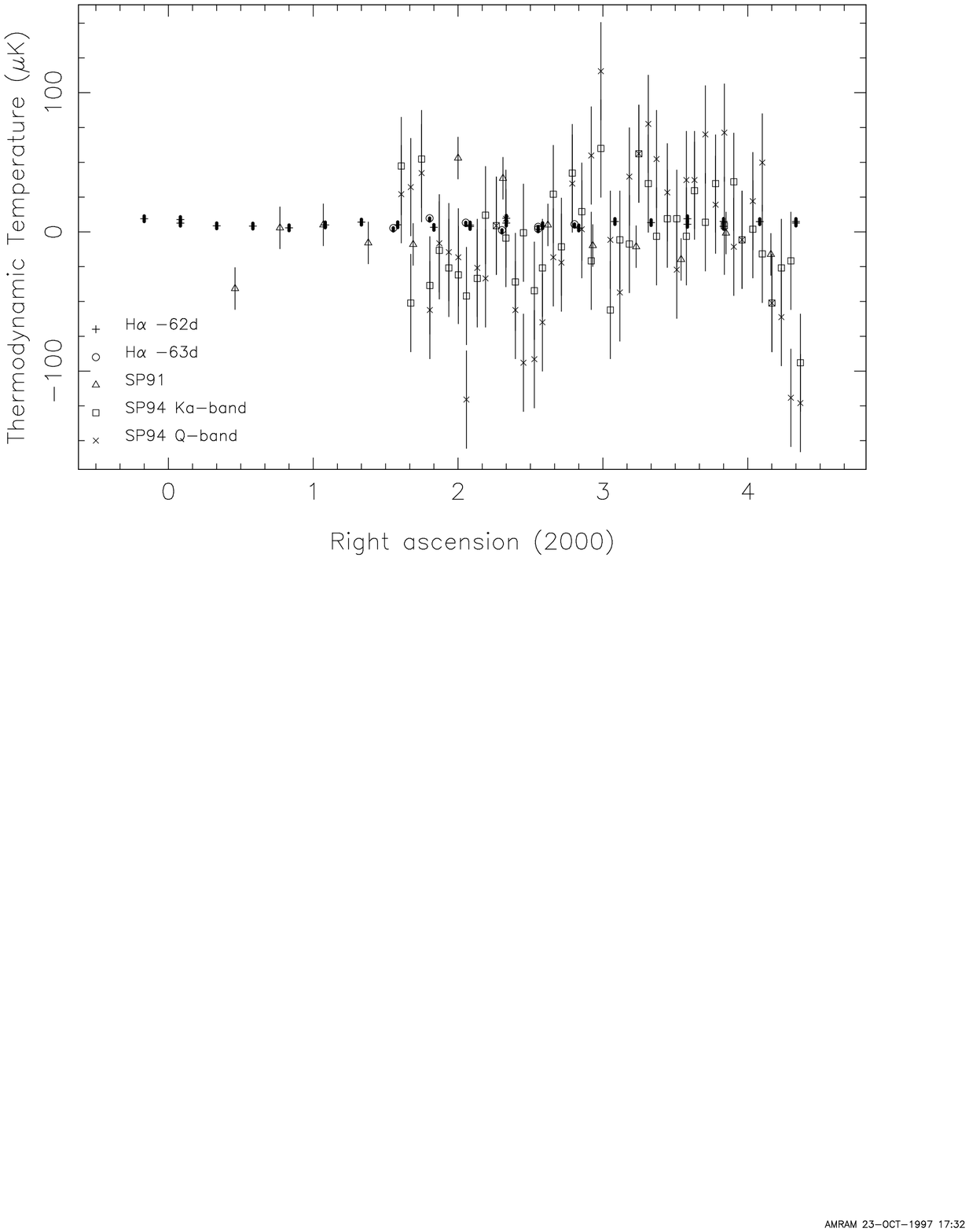,height=2.0in}
\caption{
LEFT: Solid error bars: H $\alpha$ intensities measured at
declination $-62^{\circ}$. Dotted error bars: H $\alpha$ intensities
measured at declination $-63^{\circ}$. 
RIGHT: Comparison of the South Pole
results with the free-free emission signal deduced 
from our H$\alpha$ observations.
}
\end{figure}

Fig. 2 suggests that the overall variations are faint and
that the galactic H$\alpha$ emission varies smoothly along the two bands of
sky we observed. For a gas with $T_4=1$, the magnitude of the implied
temperature variations at CMB frequencies is much smaller than observed in
the South Pole data sets.  A visual comparison is made in the 
right--hand panel of Figure 2, where we compare the {\em total}
H$\alpha$ intensity with the CMB {\em differences}.

\section{Summary}

To summarize the current status of our understanding of Galactic free--free
emission, we would say that although there is {\em no indication} of 
fluctuations large enough to pose serious difficulties for CMB
observations, the observational constraints remain weak.
A critical interpretation of the results in the tables would 
be that the limit on large scale free--free fluctuations is
the same order as the CMB amplitude on these same scales (at $~40$ GHz).
On smaller scales, observations in 
$H\alpha$ have not turned up any signs of large amplitude variations,
but those based on high resolution spectrographs are few and 
cover only a small percentage of the 
sky. There does appear to be a dust/free--free correlation 
on all angular scales, but there
is room, and perhaps tentative indications of, an equally important
non--correlated component (question \#1 posed in the introduction).  
And then there is the puzzeling
result from Leitch et al. (1997), perhaps pointing
to a hot phase of the ISM which could, due to lack of sensitivity,
escape many of the present $H\alpha$ limits (question \#2 posed in
the introduction).  

%
	
	Obviously, CMB observations at higher frequencies, where
much of the effort is now being concentrated, will suffer less
from any possible free--free contamination, and the many
of the next generation CMB experiments have a wide spectral
coverage to aid the removal of foregrounds.  Even given the above
critical viewpoint, it would be a surprise to discover 
free--free emission presenting an important difficulty for
all planned CMB experiments, at least in terms of measuring the variance,
or power spectrum, from the early Universe.  Foregrounds will,
however, be much more important for higher order statistics
looking for non--gaussian signatures.  In such cases, the
non--gaussian foregrounds will have to be removed to high
precision.   A sensitive, high spectral resolution 
H$\alpha$ survey of the entire sky would be of great value
in the context of the above considerations.

	We have also reported some results from our recent H$\alpha$
observations taken with a Fabry--Perot system at La Silla. The data cover
two bands along which the South Pole data sets show significant fluctuations
in CMB brightness. The observed H$\alpha$ fluctuations indicate that the CMB
results in this region are not significantly contaminated by free--free 
emission (assuming a gas with $T_4=1$).

\section{Acknowledgments}

We would like to thank M. Marcelin, D. Valls--Gabaud and A. Blanchard
for letting us present results based on our joint observations; 
the work presented here is the result of many
long discussions together.  We also thank G. Smoot for encouragement 
and many helpful conversations. A thanks also goes to the organizers for an
extremely enjoyable and interesting meeting.


\section*{References}

\begin{thebibliography}{99}
\bibitem{amram91} 
{\small P. Amram, J. Boulesteix, Y.M. Georgelin, Y.P. Georgelin,
A. Laval, Le Coarer E., M. Marcelin, M. Rosado, The Messenger 64, 44 (1991).}

\bibitem{Banday97}  
{\small A.J. Banday, K.M. G\'{o}rski, C.L. Bennett, G. Hinshaw, A.
Kogut, C. Lineweaver, G.F. Smoot \& L. Tenorio, ApJ 475, 393 (1997)}

\bibitem{caplan}  
{\small J. Caplan, L. Deharveng, A\&A Suppl. Ser., 62, 63 (1985)}

\bibitem{gaustad96}  {\small J.E. Gaustad, P.R. McCullough and D. Van Buren,
PASP 108, 351 (1996) }

\bibitem{gautier92}  {\small T.N. Gautier, F. Boulanger, M. P\'{e}rault and
J.L. Puget, AJ 103, 1313 (1992) }

\bibitem{Gundersen95}  
{\small J.O. Gundersen, M. Lim, J. Staren et al., ApJ 443, L57 (1995)}

\bibitem{kogut97}  {\small A. Kogut, AJ 114, 1127 (1997) }

\bibitem{kogut96a}  {\small A. Kogut , A.J. Banday, C.L. Bennett, K.M.
G\'{o}rski, G. Hinshaw and W.T. Reach, ApJ 460, 1 (1996a) }

\bibitem{kogut96b}  {\small A. Kogut, A.J. Banday, C.L. Bennett, K.M.
G\'{o}rski, G. Hinshaw, G.F. Smoot, and E.L. Wright, ApJ 464, L5 (1996b) }

\bibitem{leitch}  {\small E.M. Leitch, A.C.S. Readhead, T.J. Pearson and
S.T. Myers, ApJ 486, L23 ( 1997) }

\bibitem{Marcelin98}  
{\small M. Marcelin, P. Amram, J.G. Bartlett, D. Valls--Gabaud
\& A. Blanchard, submitted to A\&A (1998)}

\bibitem{Nossal}  {\small Nossal, PhD. thesis, University of Wisconsin,
Madison (1994) }

\bibitem{sasdust}  {\small A. de Oliveira--Costa, A. Kogut, M.J. Devlin, C.
B. Netterfield, L.A. Page and E.J. Wollack, ApJ 482, L17 (1997) }


\bibitem{reynolds90}  {\small R.J. Reynolds in ``The Galactic and
Extragalactic Background Radiation'', Eds. S. Bowyer \& C. Leinert, 157
(1990) }

\bibitem{reynolds92}  {\small R.J. Reynolds, ApJ 392, L35 (1992) }

\bibitem{reynolds98}  {\small R.J. Reynolds, S.L. Tufte, L.M. haffner, K.
Jaehnig and J.W. Percival to be published in Publications of the
Astronomical Society of Australia (can be found on the WhaM web page) (1998) 
}

\bibitem{Schuster93}  
{\small J. Schuster, T. Gaier, J. Gundersen, P. Meinhold, T.
Koch, M. Seiffert, C.A. Wuensche \& P. Lubin, ApJ 412, L47 (1993)}

\bibitem{simonetti96}  
{\small J.H. Simonetti, B. Dennison and Topasna, ApJ 458, L1
(1996) }

\bibitem{smoot98}  {\small G.F. Smoot, astro--ph/9801121 (1998) }

\bibitem{wright98} {\small E.L. Wright, ApJ 496, 1 (1998)}

\bibitem{valls98}  {\small D. Valls--Gabaud, PASA, 15, in press (1998) }

\bibitem{Yelle98}  {\small R.V. Yelle \& F.L. Roesler, J. Geophys. Res.
90,7568 (1985) }

\end{thebibliography}
\end{document}